\documentclass[preprintnumbers,amsmath,amssymb,prd]{revtex4}
\usepackage{graphicx}

\def\a{\alpha}
\def\b{\beta}

\def\d{\delta}
\def\e{\eta}

\def\l{\lambda}

\def\m{\mu}
\def\n{\nu}
\def\r{\rho}
\def\o{\omega}
\def\s{\sigma}

\def\p{\pi}
\def\e{\varepsilon}

\def\be{\begin{equation}}
\def\ee{\end{equation}}
\def\beq{\begin{eqnarray}}
\def\eeq{\end{eqnarray}}

\def\cal{\mathcal}

\def\RR{{\mathbb{R}}}
\renewcommand{\theequation}{\thesection.\arabic{equation}}

\begin{document}
\preprint{}

\title{Duality in linearized gravity}

\author{Marc Henneaux}
\affiliation{Physique Th\'eorique et Math\'ematique and
International Solvay Institutes, Universit\'e Libre de Bruxelles,
Campus Plaine C.P. 231, B--1050 Bruxelles, Belgium, \\ and
\\Centro de Estudios Cient\'{\i}ficos (CECS), Valdivia, Chile.}
\author{Claudio Teitelboim}
\affiliation{{Centro de Estudios Cient\'{\i}ficos (CECS),
Valdivia, Chile.}}
\begin{abstract}
We show that duality transformations of linearized gravity in four
dimensions, i.e., rotations of the linearized Riemann tensor and
its dual into each other, can be extended to the dynamical fields
of the theory so as to be symmetries of the action and not just
symmetries of the equations of motion. Our approach relies on the
introduction of two ``superpotentials", one for the spatial
components of the spin-2 field and the other for their canonically
conjugate momenta. These superpotentials are two-index, symmetric
tensors.  They can be taken to be the basic dynamical fields and
appear locally in the action. They are simply rotated into each
other under duality. In terms of the superpotentials, the
canonical generator of duality rotations is found to have a
Chern-Simons like structure, as in the Maxwell case.
\end{abstract}

\maketitle

\section{Introduction}
\setcounter{equation}{0} Duality is a fascinating symmetry which
keeps appearing in many contexts. Originally developed for
electromagnetism, where duality invariance of the Maxwell equations
leads to the introduction of magnetic sources and the quantization
of electric charges \cite{Dirac}, duality has been at the origin of
many remarkable developments in Yang-Mills theory \cite{'tHP}. The
generalization of duality to extended objects and $p$-form gauge
fields was carried out in \cite{Teitel,Nepo}.  More recently,
duality has revolutionized the understanding of string theory by
providing non-perturbative insight.  These latter developments
indicate that duality should play a central role in gravitation
theory as well (for a recent review, see \cite{OP}).

The idea that duality should be important in gravitation theory has
in fact a long history that goes back to the recognition that the
mixed space-time components $g_{i0}$ of the metric are analogous, in
the stationary context, to the spatial components of the vector
potential of Maxwell theory (``gravitomagnetism").  This enables one
to interpret, for instance,  the Taub-NUT solution as the metric of
an object that carries both electric-type and magnetic-type
gravitational masses. More generally, duality appears among the
``hidden symmetries" of dimensionally-reduced gravitational theories
\cite{CJ}, a subject that has received recently a new impetus
\cite{W,DH}.

Understanding gravitational duality in the general, Killing vector
free situation, remains, however, a challenge which is still
unsolved to this date.  In this article, we investigate
gravitational duality in the linearized limit {\em but without
assuming the existence of spacetime symmetries} so that usual
dimensional reduction techniques are not available. Our main result
is the proof that the action of standard linearized four-dimensional
Einstein gravity is duality invariant. This implies that duality
defines a conserved Noether charge, which acts as the generator of
duality rotations through the Poisson bracket (classically) or the
commutator (quantum-mechanically).  The existence of a Noether
generator would not hold if duality was a mere equations-of-motion
symmetry as it is sometimes assumed.  Furthermore,
duality-invariance is clearly independent (at the linearized level)
from dimensional reduction since it holds for arbitrary field
configurations.

Our approach relies on the introduction of two local
``superpotentials", which are two-index, symmetric tensors. In terms
of these superpotentials, the linearized Einstein action takes a
form similar to that of the Maxwell action.  One can then follow the
approach of \cite{DT}, where duality-invariance of the Maxwell
action was established.  The Noether charge exhibits an interesting
Chern-Simons like structure when expressed in terms of the
superpotentials.

\subsection{$SO(2)$ duality-symmetry of spin-2 free field equations
in four spacetime dimensions}

The linearized Riemann tensor $R_{\l \m \r \s}= - R_{\m \l \r \s}
= - R_{\l \m \s \r}$ fulfills the following identities, \beq
&& R_{\l [\m \r \s]} = 0, \label{I2}\\
&& R_{\l \m [\r \s, \a]} = 0 \label{I3}.\eeq It follows from
(\ref{I2}) that $R_{\l \m \r \s}$ is symmetric for the exchange of
the pairs $(\l \; \m)$ and $(\r \; \s)$, $R_{\l \m \r \s} = R_{\r
\s \l \m}$.  Moreover, the identities (\ref{I2}) and (\ref{I3})
imply the familiar fact that there exists a symmetric tensor gauge
field $h_{\m \n} = h_{\n \m}$ of which $R_{\l \m \r \s}$ is the
curvature, \be R_{\l \m \r \s} =
\partial_{[\l} \, h_{\m][\s,\,\r]}. \label{exiofh}\ee
In the absence of sources, the linearized Einstein equations take
the form \be R_{\m \n} = 0 , \label{E1}\ee where $R_{\m \n}$ is
the linearized Ricci tensor. It follows that the dual $S_{\l \m \r
\s} = - S_{\m \l \r \s} = - S_{\l \m \s \r}$ of the curvature,
defined by \be S_{\l \m \r \s} = \frac{1}{2} \, \epsilon_{\l \m \a
\b} R^{\a \b}_{\; \; \; \; \r \s} \ee enjoys
also the properties  \beq && S_{\l [\m \r \s]} = 0, \label{Ia2}\\
&& S_{\l \m [\r \s, \a]} = 0 \label{Ia3}\eeq (implying the
existence of a dual potential) and \be S_{\m \n} = 0. \label{E2}
\ee  Our conventions are as follows: the Minkowskian metric is
$\eta_{\m \n} = \hbox{diag}(-1, 1, 1, 1)$ while $\e^{0123} = 1 = -
\e_{0123}$.  Indices are lowered, raised and contracted with the
Minkowskian metric, e.g., $ R_{\m \n} = R_{\a \m \b \n} \,
\eta^{\a\b}$ and $S_{\m \n} = S_{\a \m \b \n} \, \eta^{\a\b}$.
Square brackets $[ \; \; ]$ denote antisymmetrization with
strength one, e.g., $F_{[\l \m]} = (1/2)(F_{\l \m} - F_{\m \l})$.

Comparing (\ref{I2}), (\ref{I3}) and (\ref{E1}) with (\ref{Ia2}),
(\ref{Ia3}) and (\ref{E2}) shows that the equations of the vacuum
linearized Einstein theory are invariant under duality
transformations, in which the curvature and its dual are rotated
into each other, \beq && R'_{\l \m \r \s} = \cos \a
\,R_{\l \m \r \s} + \sin \a \, S_{\l \m \r \s}, \label{dualitya}\\
&& S'_{\l \m \r \s} = - \sin \a \,R_{\l \m \r \s} + \cos \a \,
S_{\l \m \r \s} \label{dualityb}\eeq [Actually, the equations
remain invariant if we replace the rotation in (\ref{dualitya}),
(\ref{dualityb}) by a general invertible matrix. However, as we
will see in what follows, only rotations will leave the action
invariant, just as for electromagnetism.]

It is useful to rewrite the duality transformations in terms of
the electric and magnetic components of the Weyl tensor (which
coincides on-shell with the Riemann tensor). One defines \be
{\mathcal E}_{mn} = R_{0m0n} , \; \; \; {\mathcal B}_{mn} = -
\frac{1}{2} \, \epsilon_{npq}\, R_{0m}^{\; \; \; \; pq} \ee The
electric and magnetic tensors ${\mathcal E}_{mn}$ and ${\mathcal
B}_{mn}$ are both traceless and symmetric on-shell. Thus, they
have 5 independent components each, corresponding to the 10
independent components of the Weyl tensor. It is easy to verify
that the transformations (\ref{dualitya}) and (\ref{dualityb}) are
equivalent to \beq && {\mathcal E}'_{mn} = \cos \a \,{\mathcal
E}_{mn} + \sin \a \, {\mathcal B}_{mn}, \label{dualityaa}\\ &&
{\mathcal B}'_{mn} = - \sin \a \,{\mathcal E}_{mn} + \cos \a \,
{\mathcal B}_{mn} \label{dualitybb}\eeq  when the equations of
motion hold.

\subsection{Is duality a symmetry of the action?}
The question investigated in this paper is: do the duality
rotations (\ref{dualitya}) and (\ref{dualityb}) define symmetries
of the action - and thus of the theory?

In \cite{DT}, a similar question was asked for the Maxwell theory.
It was shown that duality rotations of the field strength $F_{\m
\n}$ into its dual $\!^*F_{\m \n}$ do define symmetries of the
Maxwell action. This might seem surprising at first sight since
the Maxwell Lagrangian $\sim ({\mathbf E}^2 - {\mathbf B}^2)$ is
not invariant under the (Euclidean) rotations ${\mathbf E}' = \cos
\a \, {\mathbf E} + \sin \a \, {\mathbf B}$, ${\mathbf B}' = -
\sin \a \, {\mathbf E} + \cos \a \, {\mathbf B}$. As explained in
\cite{DT}, this computation (i.e., evaluating the variation of the
Lagrangian under the duality rotations of the curvatures just
written) is meaningless, however, because the dynamical variables
in the action principle are the components of the vector potential
$A_\m$, and not the components of the field strength. Thus, in
order to investigate the invariance of the action, one must first
determine the transformation rules of the vector potential
components under duality and {\it it turns out that these are such
that the variations of the curvature take a different form
off-shell than the ones written above}. When the correct
transformation rules are used, one finds that the Maxwell action
is invariant under duality rotations \cite{DT}.

We show in this paper that the same property holds for linearized
gravity, described by the Pauli-Fierz action \be S[h_{\m\n}] =
-\frac{1}{4} \int d^4x \, \left(\partial^\r h^{\m \n}\,
\partial_\r h_{\m \n} - 2 \, \partial_\m h^{\m \n}\, \partial_\r
h^{\r}_{\; \; \n}  + 2\, \partial^\m h^\r_{\; \;\r} \,
\partial^\n h_{\m \n} - \partial^\m h^\r_{\; \;\r} \, \partial_\m
h^\s_{\; \;\s}\right).\label{PF} \ee By ``lifting" the
transformations (\ref{dualitya}) and (\ref{dualityb}) to the
fields $h_{\m \n}$, we are able to prove that the action is
duality-invariant. We also compute the corresponding conserved
charge.

The proof of duality-invariance rests on the introduction of two
spatial superpotentials, leading to a formulation analogous to the
double potential formulation of electromagnetism of Refs
\cite{DT,DGHT}. In terms of these superpotentials, manifest
duality invariance is achieved at the cost of manifest Lorentz
invariance.

\section{Superpotentials}
\setcounter{equation}{0}

\subsection{Hamiltonian form of the action}
As in \cite{DT}, we work in the Hamiltonian formalism. Any
symmetry of the Hamiltonian action is a symmetry of the original
second order action when the momenta (which can be viewed as
auxiliary fields) are eliminated through their own equations of
motion (see concluding section below).

When written in Hamiltonian form, the Pauli-Fierz action
(\ref{PF}) becomes \be S[h_{mn}, \p^{mn}, n, n_m] = \int dt \left[
\int d^3 x \, \p^{mn} \dot{h}_{mn} - H - \int d^3 x \, \left(n
{\cal H} + n_m {\cal H}^m \right)\right] \label{Haction} \ee where
$\p^{mn}$ are the conjugate momenta to the spatial components
$h_{mn}$ of the spin-2 field, while $n$ and $n_m$ are respectively
the linearized lapse and (minus 2 times) the linearized shift. The
Hamiltonian $H$ reads \be H = \int d^3x \left[\p^{mn} \p_{mn}-
\frac{1}{2} \, \p^2 + \frac{1}{4} \,
\partial^r h^{m n}\,
\partial_r h_{m n} - \frac{1}{2} \, \partial_m h^{m n}\, \partial_r
h^{r}_{\; \; n} + \frac{1}{2} \,
\partial^m h \,
\partial^n h_{m n} - \frac{1}{4} \, \partial^m h \, \partial_m
h \right] \ee where $h \equiv h^{m}_{\; \; m} $ is the trace of
the spatial $h_{mn}$ and $\p \equiv \p^{m}_{\; \; m}$ is the trace
of $\p^{mn}$. The constraints, obtained by varying the action with
respect to the lapse and the shift, are ${\cal H} =0$ and ${\cal
H}^m =0$ with \beq && {\cal H} =
\partial^{m}
\partial^n h_{mn} - \Delta h
\\ && {\cal H}^m = \p^{mn}_{\; \; \; \; \; , n}\eeq where $\Delta
\equiv \partial^m \partial_m$ is the spatial Laplacian.

\subsection{Solution of the momentum constraint - First Superpotential}
In order to exhibit the duality symmetry, we solve the
constraints. This can be achieved while maintaining locality of
the action principle by introducing ``superpotentials".

The general solution of the constraint ${\cal H}^m =0$ is (see
Appendix) \be \p^{mn}=
\partial_p
\partial_r U^{mpnr} \label{solforp0}\ee where the tensor $U^{mpnr}$
has the antisymmetry properties $U^{mpnr} = - U^{pmnr} = U^{nrmp}
= - U^{mprn}$. By dualizing $U^{mpnr}$ in terms of a symmetric
tensor $P_{qs} = P_{sq}$, \be U^{mpnr} = \epsilon^{mpq}
\epsilon^{nrs} P_{qs} ,\ee this expression can be rewritten \beq
\p^{mn} &=& \epsilon^{mpq} \epsilon^{nrs}
\partial_p
\partial_r P_{qs} \\&=& \d^{mn} (\Delta P - \partial^r \partial^s P_{rs}) +
\partial^m \partial^r P^n_{\; \; \; r} + \partial^n \partial^r P^m_{\; \; \; r}
- \partial^m \partial^n P - \Delta P^{mn} \label{solforp}\eeq
where $P$ is the trace of $P_{mn}$.  We shall call the symmetric
tensor $P_{mn}$, which is equivalent to $U^{mpnr}$,
``superpotential" (for the momenta).

Given $\p^{mn}$, the superpotential $P_{mn}$ is determined up to
$$ P_{mn} \longrightarrow P_{mn} +\partial_m \xi_n +\partial_n
\xi_m.
$$ A quick way to see this is to observe that  $\p^{mn}$
may be viewed as the Einstein tensor of $P_{mn}$ regarded as a
metric. In three dimensions, the Einstein tensor completely
determines the Riemann tensor, and hence $\p^{mn}$ determines
$P_{mn}$ up to a gauge transformation $\partial_m \xi_n
+\partial_n \xi_m$.  A particular solution is \be P_{mn} = -
\Delta^{-1} \p_{mn} + \delta_{mn} \Delta^{-1} \p \label{P}\ee One
easily verifies that (\ref{P}), when inserted in (\ref{solforp}),
reproduces $\p^{mn}$ satisfying the constraint ${\cal H}^m =0$.

One may use $P_{mn}$ instead of $\pi^{mn}$ as fundamental field in
the action principle.  When this is done, the momentum constraint
and its Lagrange multiplier $n_m$ drop out from the action
principle because the constraint is identically satisfied.
Although the expression of $\pi^{mn}$ is local in terms of
$P_{mn}$ (which is what matters for the locality of the action
expressed in terms of $P_{mn}$), the inverse transformation is
non-local.

The momenta $\p^{mn}$ are not gauge invariant but transform as
$\p^{mn}  \longrightarrow \p^{mn} - \partial^m \partial^n \xi +
\d^{mn} \Delta \xi $ under the transformation generated by the
Hamiltonian constraint.  This transformation is simply generated
by a conformal change of the superpotential $P_{mn}$, $\d P_{mn} =
\d_{mn} \xi$.  Hence, the total ambiguity in $P_{mn}$ is \be
P_{mn} \longrightarrow P_{mn} +\partial_m \xi_n +\partial_n \xi_m
+ \d_{mn} \xi \label{ambP}\ee  The transformations (\ref{ambP})
appear as gauge transformations of the formulation in which
$P_{mn}$ is regarded as the fundamental field.

\subsection{Solution of the Hamiltonian constraint - Second
Superpotential} Similarly, one can also solve the ``Hamiltonian
constraint" ${\cal H} =0$ in terms of a symmetric superpotential
$\Phi_{mn} = \Phi_{nm}$ and a vector $u_m$ as \be h_{mn} =
\epsilon_{mrs} \,
\partial^r \Phi^{s}_{\; \; \; n} + \epsilon_{nrs} \, \partial^r
\Phi^{s}_{\; \; \; m} + \partial_m u_n + \partial_n u_m
\label{solforh}\ee (see Appendix).

Given $h_{mn}$ up to a gauge transformation, there is some
ambiguity in the superpotential $\Phi_{mn}$, which reads exactly
as in (\ref{ambP}), \be \Phi_{mn} \longrightarrow \Phi_{mn}
 +\partial_m \eta_n +\partial_n \eta_m + \d_{mn} \eta \label{ambphi}\ee
This is also shown in the Appendix.

One can express $ \Phi_{mn}$ non-locally in terms of the metric. A
particular solution is \be \Phi_{mn} = -\frac{1}{4} \, \Delta^{-1}
\left(\e_{mrs} \, \partial^r h^s_{\; \; n} + \e_{nrs} \,
\partial^r h^s_{\; \; m} \right) \label{solforphi}\ee with \be u_m
= \frac{1}{4}\, \Delta^{-1} \left(3\,  \partial^p h_{pm} -
\partial_m h \right) . \ee We leave it to the reader to verify
that this expression leads back to an $h_{mn}$ obeying the
constraint ${\cal H} =0$ when inserted in (\ref{solforh}). When
$\Phi_{mn}$ is used as fundamental field instead of $h_{mn}$,
there is no constraint left and (\ref{ambphi}) appears as a gauge
transformation.

Note that the first order constraint ${\cal H}^m =0$ yields an
expression for the momenta that involves two derivatives of the
superpotential $P_{mn}$, while the second order constraint ${\cal
H} =0$ yields an expression for $h_{mn}$ that involves only one
derivative of $\Phi_{mn}$ and $u_m$. Accordingly, the Hamiltonian
is a homogeneous polynomial quadratic in the second derivatives of
both superpotentials.

\section{Duality transformations in terms of superpotentials}
\setcounter{equation}{0}

It turns out that the duality rotations are simply $SO(2)$
rotations of the superpotentials into each other, \beq && P'^{mn}
= \cos \a \,P^{mn} + \sin \a \, \Phi^{mn}, \label{dualityc}\\ &&
\Phi'^{mn} = - \sin \a \,P^{mn} + \cos \a \, \Phi^{mn}
\label{dualityd}\eeq

To verify this assertion, we check that (\ref{dualityc}) and
(\ref{dualityd}) imply (\ref{dualityaa}) and (\ref{dualitybb})
on-shell.  To this end, we observe that on-shell, ${\mathcal
E}_{mn} = \!^{(3)} R_{mn}$.  Substituting the expression for
$h_{mn}$ in terms of the superpotential $\Phi_{mn}$, one gets \be
{\mathcal E}_{mn} = - \e_{mpq} \, \partial_n \partial^k
\partial^p \Phi^q_{\; \; k} - \e_{npq} \, \partial_m \partial^k
\partial^p \Phi^q_{\; \; k}  +
\e_{mpq} \,
\partial^p \Delta \Phi^q_{\; \; n} + \e_{npq} \, \partial^p \Delta
\Phi^q_{\; \; m}\label{EE}\ee Similarly, one finds that
$${\mathcal B}_{mn} = - \e_{mpq} \,
\partial^p \left(\p_n^{\; \; q} - \frac{1}{2} \d_n^{\; q} \right) - \e_{npq}
\, \partial^p \left(\p_m^{\; \; q} - \frac{1}{2} \d_m^{\;
q}\right)$$ when the equations of motion hold, which easily yields
\be {\mathcal B}_{mn} = - \e_{mpq} \, \partial_n \partial^k
\partial^p P^q_{\; \; k} - \e_{npq} \, \partial_m \partial^k
\partial^p P^q_{\; \; k} + \e_{mpq}
\, \partial^p \Delta P^q_{\; \; n} + \e_{npq} \, \partial^p \Delta
P^q_{\; \; m}\ee  This expression for the magnetic components of
the Weyl tensor is the same as the expression (\ref{EE}) for the
electric components of the Weyl tensor, with $P_{mn}$ replacing
$\Phi_{mn}$.  Since these expressions are linear in $P_{mn}$ and
$\Phi_{mn}$, rotations of $P_{mn}$ and $\Phi_{mn}$ into each other
induce indeed the electric-magnetic duality rotations.

\section{Duality invariance of the action}
\setcounter{equation}{0}
\subsection{Duality invariance of the Hamiltonian}
We now insert the above expressions (\ref{solforp}) and
(\ref{solforh}) in the Hamiltonian. Tedious but straightforward
computations give for the kinetic energy density (up to total
derivatives that are being dropped) \be \p^{ij} \p_{ij} -
\frac{1}{2} \p^2 = \Delta P_{ij} \, \Delta P^{ij} + \frac{1}{2}\,
(\partial^k
\partial^m P_{km})^2 +
\partial^k
\partial^m P_{km}\, \Delta P  - 2\, \partial_m
\partial_i P^{ij}\,
\partial^{m}\partial^k P_{kj} -\frac{1}{2} \, (\Delta P)^2 \ee
Similarly, the potential energy density becomes (up to total
derivatives) \beq && \frac{1}{4} \,
\partial^r h^{m n}\,
\partial_r h_{m n} - \frac{1}{2} \, \partial_m h^{m n}\, \partial_r
h^{r}_{\; \; n} + \frac{1}{2} \,
\partial^m h \,
\partial^n h_{m n} - \frac{1}{4} \, \partial^m h \, \partial_m
h \nonumber \\ && \hspace{1.5cm} = \; \Delta \Phi_{ij} \, \Delta
\Phi^{ij} + \frac{1}{2}\, (\partial^k \partial^m \Phi_{km})^2 +
\partial^k
\partial^m \Phi_{km}\, \Delta \Phi - 2\, \partial_m
\partial_i \Phi^{ij}\,
\partial^{m}\partial^k \Phi_{kj} -\frac{1}{2} \, (\Delta \Phi)^2 \eeq
[This computation is simplified once it is recalled that the
Hamiltonian is gauge-invariant. One may thus set $u_m = 0$ in the
expression (\ref{solforh}) when evaluating $H$.]

Because the kinetic and potential energies take exactly the same
form in terms of their respective superpotentials, one sees that
the Hamiltonian is invariant under $SO(2)$-rotations in the plane
of $P^{mn}$ and $\Phi^{mn}$, i.e., the Hamiltonian is duality
invariant.

Note that although not manifestly so, the Hamiltonian is positive.
This can be be seen for instance by Fourier-transforming $P^{mn}$
and $\Phi^{mn}$. Assuming a single Fourier mode propagating in the
third spatial direction, one gets for the energy density in
$k$-space \be  (k^3)^4 \left[2  (P^{11} - P^{22})^2 + 8 (P^{12})^2
+ 2 (\Phi^{11} - \Phi^{22})^2 + 8 (\Phi^{12})^2 \right]\ee

\subsection{Duality invariance of the kinetic term}
The invariance of the kinetic term $\pi \, \dot{h}$ can also be
checked easily. Injecting the expressions (\ref{solforp}) and
(\ref{solforh}) into $\p^{mn} \, \dot{h}_{mn}$, one gets \be \int
dt \, d^3x \, \p^{mn} \, \dot{h}_{mn} = 2  \int dt \, d^3x \,
\epsilon^{mrs} \, \left(
\partial^p
\partial_q
\partial _r P_{ps} - \Delta \partial_r
P_{qs}\right) \dot{\Phi}^q_{\; \; m}\ee  Because this expression
changes sign (up to a total derivative) under the exchange of
$P^{mn}$ with $\phi^{mn}$, it is invariant under the rotations
(\ref{dualityc}) and (\ref{dualityd}) (up to a total derivative).
This ends the proof of the duality-invariance of the free massless
spin-2 theory in four dimensions.

\subsection{$SO(2)$-vector notations}
By introducing $SO(2)$-vector notations and adding a total
derivative to make the kinetic term strictly antisymmetric under
the exchange of the superpotentials, one may rewrite the free
spin-2 action -- with the superpotentials as basic dynamical
fields -- as \be S[Z_a^{\; mn}] = \int dt \left[ \int d^3x \,
\epsilon^{ab}\epsilon^{mrs} \, \left(
\partial^p
\partial_q
\partial _r Z_{aps} - \Delta \partial_r
Z_{aqs}\right) \dot{Z}^{\; q}_{b\; m} - H \right] \ee with \be
(Z_a^{\; mn}) = (P^{mn}, \Phi^{mn}), \; \; \; \; a,b = 1,2 \ee and
\be H = \int d^3x \, \delta^{ab} \left(\Delta Z_{aij} \, \Delta
Z_b^{\;ij} + \frac{1}{2}\,
\partial^k
\partial^m Z_{a km}\partial^q
\partial^n Z_{b qn} +
\partial^k
\partial^m Z_{akm}\, \Delta Z_b - 2\,
\partial_m
\partial_i Z_a^{\;ij}\,
\partial^{m}\partial^k Z_{bkj} -\frac{1}{2} \, \Delta Z_a
\Delta Z_b \right)  \ee with $Z_a \equiv Z_{am}^{\; \; \; \; m}$.

This expression is manifestly duality invariant because the
tensors $\epsilon^{ab}$ and $\delta^{ab}$ are $SO(2)$-invariant.
It should be compared with the analogous expression for the
Maxwell action considered in \cite{DGHT}.  We also see that linear
transformations that leave the action invariant must preserve both
$\epsilon^{ab}$ and $\delta^{ab}$ and thus necessarily belong to
$SO(2)$.

\section{Duality generator}
\setcounter{equation}{0}  The conserved charge that generates
infinitesimal duality rotations is found from the Noether theorem
to be \be Q = \frac{1}{2}\int d^3 x \, \e^{mrs}\,
\left[(\partial^p
\partial_q \partial_r P_{ps} - \Delta \partial_r P_{qs}) P^q_{\;
\; m} -(\partial^p
\partial_q \partial_r \Phi_{ps} - \Delta \partial_r \Phi_{qs}) \Phi^q_{\;
\; m} \right] \label{Q}\ee  It is invariant under the respective
gauge transformations (\ref{ambP}) and (\ref{ambphi}) of $P_{mn}$
and $\Phi_{mn}$.

One may rewrite the conserved charge more suggestively by
introducing the curvatures and spin connections of $P_{mn}$ and
$\Phi_{mn}$.  These are defined by \beq && R(P)_{pqrs} =
\partial_{[q}P_{p][r,s]}, \; \; \; \; \o(P)^{pq}_{\; \; \; \; m} =
\partial^p P^q_{\; \; m} - \partial^q P^p_{\; \; m} \\
&& R(\Phi)_{pqrs} =
\partial_{[q}\Phi_{p][r,s]}, \; \; \; \; \o(\Phi)^{pq}_{\; \; \; \; m} =
\partial^p \Phi^q_{\; \; m} - \partial^q \Phi^p_{\; \; m} \eeq
Upon integration by parts, (\ref{Q}) becomes \be Q = \frac{1}{2}
\int d^3x \, \e^{mrs} \left( R(P)_{pqrs} \, \o(P)^{pq}_{\; \; \;
\; m} - R(\Phi)_{pqrs} \, \o(\Phi)^{pq}_{\; \; \; \; m} \right)
\ee In terms of the curvature two-forms $R(P)_{pq} = \frac{1}{2}
R(P)_{pqrs} \, dx^r \wedge dx^s$,  $R(\Phi)_{pq} = \frac{1}{2}
R(\Phi)_{pqrs} \, dx^r \wedge dx^s$ and the connection one-forms
$\o(P)^{pq} = \o(P)^{pq}_{\; \; \; \; m}\, dx^m$, $\o(\Phi)^{pq} =
\o(\Phi)^{pq}_{\; \; \; \; m} \, dx^m$, one can rewrite this
expression as \be Q = \int \left[R(P)_{pq} \wedge \o(P)^{pq} -
R(\Phi)_{pq} \wedge \o(\Phi)^{pq} \right] \ee exhibiting a
Chern-Simons structure analogous to that found in the Maxwell
case.

Under the gauge transformations $ P_{mn} \rightarrow P_{mn} +
\partial_m \xi_n + \partial_n \xi_m$, the curvature $R(P)_{pqrs}$
is invariant, while $\o(P)^{pq}_{\; \; \; \; m}$ transforms as the
gradient of the rotation parameter $\partial^p \xi^q -
\partial^q \xi^p$, $$\o(P)^{pq}_{\; \; \; \; m} \rightarrow \o(P)^{pq}_{\;
\; \; \; m} + \partial_m (\partial^p \xi^q - \partial^q
\xi^p).$$This is the transformation property of the spin
connection in linearized gravity when the local rotation gauge
freedom is fixed by the gauge condition that the triads should be
symmetric. Hence, the name ``spin connection" for $\o(P)^{pq}_{\;
\; \; \; m}$. The same properties hold for $R(\Phi)_{pqrs}$ and
$\o(\Phi)^{pq}_{\; \; \; \; m}$ under the gauge transformations
$\Phi_{mn} \rightarrow \Phi_{mn} +
\partial_m \eta_n + \partial_n \eta_m$.

\section{Conclusion and summary}
\setcounter{equation}{0}

In this paper, we have shown that duality is a symmetry not only
of the equations of motion of the free spin-2 theory but also of
the Pauli-Fierz action itself.  Hence, duality is a symmetry of
(linearized) gravity in the standard sense. This was achieved by
introducing symmetric superpotentials. In terms of these
superpotentials, the action takes a form very similar to that of
electromagnetism (compare with ``conformal gravity" analyzed in
\cite{DN}).  We have also computed the canonical generator of
duality rotations and found the same Chern-Simons structure as in
the spin-1 case.  The theory is invariant under the gauge
transformations (\ref{ambP}) and (\ref{ambphi}) of the
superpotentials, which take the form of independent linearized
diffeomorphisms and conformal transformations.  As in
electromagnetism, the price paid for achieving manifest
duality-invariance of the action is the loss of manifest
Lorentz-invariance.

We have explicitly written the duality transformation rules in
terms of the electric and magnetic superpotentials and verified
duality-invariance only for the reduced action where the
constraints have been eliminated, but this is sufficient to
establish invariance of the original Pauli-Fierz action itself.
This is in fact a standard general result, but for the sake of
completeness, we repeat the reasoning here. The argument proceeds
in two steps:
\begin{enumerate} \item First, one proves duality invariance of the
unreduced Hamiltonian action (\ref{Haction}).  From the
transformation properties of the superpotentials, one can infer
the transformation properties of $h_{mn}$ and $\pi^{mn}$ using $\d
\Phi_{mn} = P_{mn}$, $\d P_{mn} = - \Phi_{mn}$,  (\ref{solforh}),
(\ref{P}), (\ref{solforp}) and (\ref{solforphi}). One gets
(assuming $\delta u_m =0$ for simplicity, which is legitimate)
\beq \d h_{mn} &=& - \e_{mrs} \, \partial^r(\Delta^{-1} \pi^s_{\;
\; n}) -
\e_{nrs} \, \partial^r(\Delta^{-1} \pi^s_{\; \; m})\\
\d \p^{mn} &=& \frac{1}{4} \e^{nrs} \left( \partial_p \partial_r
\partial^m \Delta^{-1} h^{p}_{\; \; s} - \partial_r h^m_{\; \; s}
\right) \nonumber \\ && + \frac{1}{4} \e^{mrs}\left( \partial_p
\partial_r
\partial^n \Delta^{-1} h^{p}_{\; \; s} - \partial_r h^n_{\; \; s}
\right) \eeq These transformation rules hold on the constraint
surface and we choose to extend them off the constraint surface
using the same expressions.  From what we have shown, the
unreduced Hamiltonian action (\ref{Haction}) is invariant under
these transformations when the constraints hold, or, what is the
same, its variation under duality is a combination of the
constraints.  One may thus adjust the variations of the Lagrange
multipliers $n$ and $n_m$, which are free so far, so as to cancel
the constraint terms that appear. This is always possible since
the Lagrange multipliers multiply the constraints.  The
computation is cumbersome and will not be reproduced here. \item
Second, one eliminates the momenta $\pi^{mn}$ using their own
equations of motion, i.e., through \be \pi^{mn} = \frac{1}{2}
\left( \dot{h}^{mn} - \d^{mn} \dot{h} \right) \label{momenta} \ee
The standard general theorems on auxiliary fields guarantee that
the resulting action, which is the just the Pauli-Fierz action
(\ref{PF}), is invariant under the transformations in which the
momenta are replaced by their on-shell values (\ref{momenta}).
This gives in particular, the following duality transformation
rules for the spatial components of the metric \be \d h_{mn} = -
\frac{1}{2}\left[\e_{mrs}
\partial^r(\Delta^{-1} \dot{h}^s_{\; \; n}) + \e_{nrs}
\partial^r(\Delta^{-1} \dot{h}^s_{\; \; m})\right] \ee Note that
this expression is non-local in space but local in time.
\end{enumerate}

Although we have not carried it explicitly, we expect the
discussion of the duality properties of higher spins gauge fields
actions\cite{BB03} to proceed similarly. Much more challenging
would the understanding of how the results can be extended to the
full, non linear Einstein theory (in the same Killing vector free
context considered here). The inclusion of dynamical sources of
both electric and magnetic types in the general context is also an
intricate question.

It is well known that the (full) Einstein action dimensionally
reduced to 3 spacetime dimensions exhibits a ``hidden" $SL(2,\RR)$
symmetry, of which a $SO(2)$ subgroup acts linearly in the small
field limit and is the duality group considered above \cite{Geroch}.
We have shown that this subgroup is already a symmetry of the
Einstein action prior to dimensional reduction, at least in the
linearized theory.  The independence of the existence of duality (as
a symmetry of the action) on dimensional reduction appears to be
very suggestive.

\section*{Acknowledgments}
The work of MH is partially supported by IISN - Belgium
(convention 4.4505.86), by the ``Interuniversity Attraction Poles
Programme -- Belgian Science Policy '' and by the European
Commission RTN programme HPRN-CT-00131, in which he is associated
to K. U. Leuven. Institutional support to the Centro de Estudios
Cient\'{\i}ficos (CECS) from Empresas CMPC is gratefully
acknowledged.  CECS is a Millennium Science Institute and is
funded in part by grants from Fundaci\'on Andes and the Tinker
Foundation.

\appendix

\section{Proof of Equations (\ref{solforp0}) (\ref{solforh}) and (\ref{ambphi})}

\renewcommand{\theequation}{A.\arabic{equation}}
We supply in this appendix the proofs of results used -- but left
unproved -- in the main text.  These results have a cohomological
content in that they may be viewed as special cases of the
``Poincar\'e lemma" for a first-order differential operator $D$
fulfilling $D^3 = 0$ and generalizing to tensors characterized by
two-column Young tableaux, the familiar exterior derivative
operator $d$ ($d^2 = 0$) appropriate to antisymmetric tensors
(characterized by one-column Young tableaux) \cite{DVH1,O}.  In
that spirit, the Bianchi identity (\ref{I3}) can for instance be
written as $DR = 0$ while (\ref{exiofh}) becomes $R = D^2h$
\cite{DVH1}. We shall, however, not use the general results
established in \cite{DVH1,O} to derive (\ref{solforp0})
(\ref{solforh}) and (\ref{ambphi}) but instead, we shall follow a
more explicit approach, which is in fact rather direct in the
present case.
\subsection{Proof of Eq. (\ref{solforp0})}

We first show that (\ref{solforp0}) is the general solution of
$\p^{m n}_{\; \; \; \; , \, n} = 0$.  The standard Poincar\'e
lemma for closed $2$-forms, with $m$ viewed as a spectator index,
yields $$\p^{m n} = \partial_k M^{mnk}$$ with $M^{mnk} = -
M^{mkn}$.  The symmetry of $\p^{mn}$ for the exchange of $m$ with
$n$ implies then $\partial_k (M^{mnk} - M^{nmk}) = 0$, from which
one infers, using again the standard Poincar\'e lemma (with $m$
and $n$ regarding as an antisymmetric pair of spectator indices)
that
$$M^{mnk} - M^{nmk} = \partial_s A^{mnks} $$ with $A^{mnks} = -
A^{mnsk}= - A^{nmks}$.  This leads to Eq. (\ref{solforp0}),
$$\p^{m n} = \partial_k  \partial_s  U^{mkns}$$ with $$U^{mkns}
\equiv - \frac{1}{2} \left(A^{mkns} + A^{nsmk} \right) = -
U^{kmns} = - U^{mksn} = U^{nsmk}.$$

\subsection{Proof of Equ. (\ref{solforh})}

We now prove that the general solution of \be
\partial^{m}
\partial^n h_{mn} - \Delta h = 0 \label{eq1}\ee is
\be h_{mn} = \epsilon_{mrs}
\partial^r \Phi^{s}_{\; \; \; n} + \epsilon_{nrs} \partial^r
\Phi^{s}_{\; \; \; m} + \partial_m u_n + \partial_n u_m
\label{solsolforh}\ee where the superpotential $\Phi^{mn}$ is
symmetric, $\Phi^{mn} = \Phi^{nm}$.

To that end, we first note that that one can write $h_{mn}$ as $$
h_{mn} = j_{mn} + \partial_m v_n + \partial_n v_m $$ where
$j_{mn}$ is symmetric and traceless. The tensor $h_{mn}$ is a
solution of (\ref{eq1}) if and only if $j_{mn}$ is a solution of
\be
\partial^{m}
\partial^n j_{mn} = 0 \label{eq2}\ee This equation can be written
as $\partial^m (\partial^n j_{mn})= 0$ from which we get, using
the standard Poincar\'e lemma, $\partial^n j_{mn} = \e_{mnq} \,
\partial^n M^q $ for some $M^q$ or equivalently
$$ \partial^n (j_{mn} - \e_{mnq} \, M^q ) = 0.$$  Using again the
standard Poincar\'e lemma with $m$ a spectator index yields $$
j_{mn} - \e_{mnq} \, M^q = 2 \, \e_{nrs}\partial^r \Psi^{s}_{\; \;
\; m} $$ for some $\Psi^{s}_{\; \; \; m}$.  Taking the symmetric
part gives \be j_{mn} = \epsilon_{mrs}\,
\partial^r \Psi^{s}_{\; \; \; n} + \epsilon_{nrs} \partial^r
\Psi^{s}_{\; \; \; m}. \label{solforj1}\ee {}For arbitrary
$j_{mn}$, $\Psi^{rs}$ would not be symmetric.  However, $j_{mn}$
is not arbitrary but is traceless. This implies further conditions
on $\Psi^{rs}$. From $j^m_{\;\;m} = 0$, one gets
$\partial_{[p}\Psi_{qm]} = 0$ from which it follows that
$\psi_{[qm]} = \partial_q w_m -
\partial_m w_q$ for some $w_m$. The antisymmetric part of
$\Psi^{rs}$ contributes therefore to Eq.(\ref{solforj1}) a term of
the form $$
\partial_n (-\epsilon_{mrs} \partial^r w^s) + \partial_m
(-\epsilon_{nrs}
\partial^r w^s)$$ which can be absorbed in the vector $v_m$. When
this is done, one finds that $h_{mn}$ takes indeed the form
(\ref{solsolforh}), as announced.

\subsection{Proof of Equ. (\ref{ambphi})}

Given  $h_{mn}$, what is the ambiguity in $(\Phi_{mn}, u_m)$?  To
answer this question, we must analyze the homogeneous equation \be
0 = \epsilon_{mrs}
\partial^r \Phi^{s}_{\; \; \; n} + \epsilon_{nrs} \partial^r
\Phi^{s}_{\; \; \; m} + \partial_m u_n + \partial_n u_m \ee
Taking the trace of that equation yields $\partial^m u_m = 0$,
i.e., $u_m = \e_{mpq} \partial^p \chi^q$ for some $\chi^q$.  This
yields \be  0  = \epsilon_{mrs}
\partial^r \Psi^{s}_{\; \; \; n} + \epsilon_{nrs} \partial^r
\Psi^{s}_{\; \; \; m} \label{eqforpsi}\ee for \be \Psi_{mn} =
\Phi_{mn}+
\partial_m \chi_n +
\partial_n \chi_m \label{defpsi}\ee
Taking the divergence of (\ref{eqforpsi}) with respect to $m$
gives $\e_{npq}
\partial^p (\partial^m \Psi^q_{\; \; \; m}) = 0$, i.e. $\partial^m
\Psi^q_{\; \; \; m} =
\partial^q C$ for some $C$.  Eq. (\ref{eqforpsi}) becomes then $$
\partial^i \left(\Psi^j_{\; \; n} +  \delta^j_{\;n}  B \right) -
\partial^j \left(\Psi^i_{\; \; n} +  \delta^i_{\;n} B
\right) = 0 $$ with $ B = (1/2) (C + \Psi)$, from which one infers
$$ \Psi_{in} =
\partial_i k_n -  \delta_{in} \,  B $$ for some $k_n$.  The
symmetry of $\Psi_{in}$ implies then $k_n = \partial_n D$ for some
$D$ and thus we get finally $ \Psi_{mn} = \partial_m \partial_n D
-  \delta_{in} B $.  Combined with (\ref{defpsi}), this shows that
the ambiguity in $\Phi_{mn}$ is indeed of the form \be \Phi_{mn}
\longrightarrow \Phi_{mn}
 +\partial_m \eta_n +\partial_n \eta_m + \d_{mn} \eta  \label{A9}\ee
 as in (\ref{ambphi}).  The ambiguity in $u_m$, is
 $u_m \rightarrow u_m - \e_{mrs}\partial^r \eta^s $.
 If one gives $h_{mn}$ only up to a gauge
 transformation, one finds that $\Phi_{mn}$ is still given
 up to (\ref{A9}) while $u_m$ can be shifted at will,  $ u_m
 \rightarrow u_m + \xi_m$.

\end{document}